\documentclass[aps,apl,twocolumn,superscriptaddress,floatfix]{revtex4}
\usepackage{amsfonts}
\usepackage{amsmath}
\usepackage{amssymb}
\usepackage{graphicx}

\begin{document}
\title{First harmonic measurements of the spin Seebeck effect}
\author{Yizhang Chen}
\thanks{These two authors contributed equally}
\affiliation{Center for Quantum Phenomena, Department of Physics, New York University, New York, New York 10003, USA}
\author{Debangsu Roy}
\thanks{These two authors contributed equally}
\affiliation{Center for Quantum Phenomena, Department of Physics, New York University, New York, New York 10003, USA}
\author{Egecan Cogulu}
\affiliation{Center for Quantum Phenomena, Department of Physics, New York University, New York, New York 10003, USA}
\author{Houchen Chang}
\affiliation{Department of Physics, Colorado State University, Fort Collins, Colorado 80523, USA}
\author{Mingzhong Wu}
\affiliation{Department of Physics, Colorado State University, Fort Collins, Colorado 80523, USA}
\author{Andrew D. Kent}
\email{andy.kent@nyu.edu}
\affiliation{Center for Quantum Phenomena, Department of Physics, New York University, New York, New York 10003, USA}
\date{\today}


\begin{abstract}
We present measurements of the spin Seebeck effect (SSE) by a technique that combines alternating currents (AC) and direct currents (DC). The method is applied to a ferrimagnetic insulator/heavy metal bilayer, Y$_3$Fe$_5$O$_{12}$(YIG)/Pt. Typically, SSE measurements use an AC current to produce an alternating temperature gradient and measure the voltage generated by the inverse spin-Hall effect in the heavy metal at twice the AC frequency. Here we show that when Joule heating is associated with AC and DC bias currents, the SSE response occurs at the frequency of the AC current drive and can be larger than the second harmonic SSE response. We compare the first and second harmonic responses and show that they are consistent with the SSE. The field dependence of the voltage response is used to characterize the damping-like and field-like torques. This method can be used to explore nonlinear thermoelectric effects and spin dynamics induced by temperature gradients.

\end{abstract}
\maketitle

A central theme in spintronics is the interconversion of charge and spin currents~\cite{Brataas2012}. Recently, a focus has been on magnetic insulators where spin transport occurs through spin-wave propagation and spin currents can be generated by either spin injection~\cite{Cornelissen2018} or by thermal gradients~\cite{uchida2008observation, BarryZink}. These phenomena can be studied in simple bilayer films consisting of a ferrimagnetic (FIM) insulator, such as Y$_3$Fe$_5$O$_{12}$(YIG), and a heavy metal (HM) with large spin-orbit coupling such as Pt. Spin to charge current conversion in such bilayers occurs by the inverse spin-Hall~\cite{Dyakonov,Hirsch,Zhang} and Rashba-Edelstein effects~\cite{Rojas-Sanchez,Shen}.

Spin to charge conversion enables determination of the spin Seebeck effect (SSE)\cite{uchida,sch, Saitoh, SSE_CLChien}. A thermal gradient across the FIM film produces a spin current into a neighboring heavy metal film, resulting in a transverse charge current or a voltage across the heavy metal film in an open circuit situation. This leads to a convenient route to characterize the spin transport as well as a means to study the inverse effects, such as the spin torque on the FIM magnetization in response to spin currents associated with charge current flow in the HM. In fact, the SSE also enables detection of the FIM magnetization direction by relatively simple electrical measurements.

In this article, we present first harmonic measurements of the SSE by a technique that combines AC and DC currents in a YIG/Pt bilayer. The temperature gradient is created by Joule heating in a Pt strip and both the linear and nonlinear responses in the longitudinal and transverse voltages are determined as a function of the angle between the AC current and an in-plane external magnetic field. An analysis of the responses shows that the SSE accounts for the main component of the second harmonic voltage response, corroborating results in the literature~\cite{avcix}. However, when a DC current is present, several new features are observed. First, we detect field-induced switching of the YIG magnetization in both the first and second harmonic longitudinal voltage measurements. Second, both the first and second harmonic transverse voltages show a step when the magnetization reverses. Interestingly, the step height in the first harmonic response has a linear dependence on DC current density and cosine dependence on in-plane field angle. Of particularly interest is that the presence of a DC current superposed on the AC current enables measurements of the SSE in the first harmonic response with an increase in signal amplitude relative to the second harmonic signal.

\begin{figure}
\includegraphics[width=0.5\textwidth]{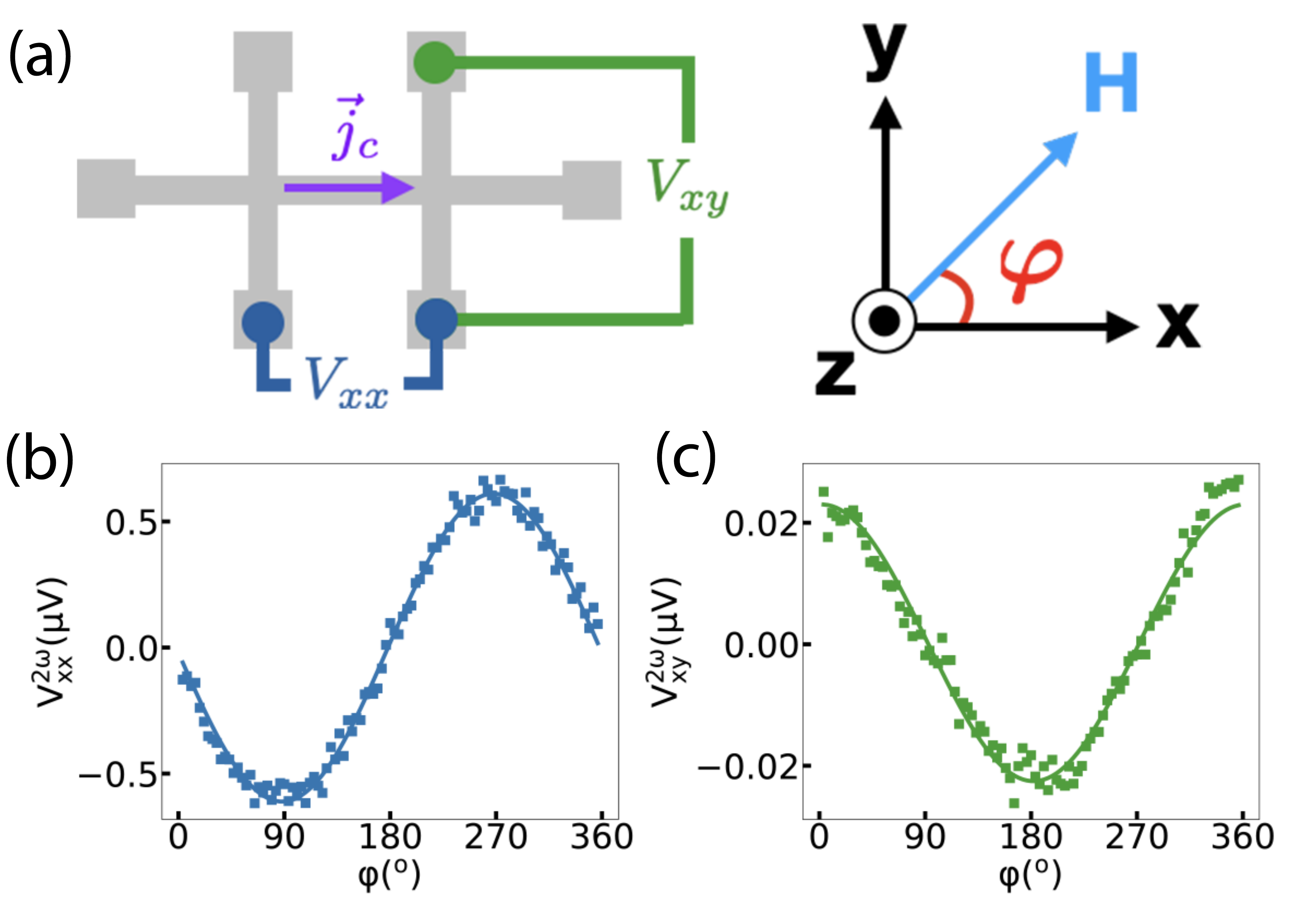}
\caption{(a) Measurement setup. $\vec{j}_c$ is the charge current density along the x direction. $V_{xx}$ and $V_{xy}$ are voltages measured in the longitudinal and transverse directions, respectively, while $\varphi$ is the angle between the applied field and the current. (b) Angular dependence of second harmonic longitudinal voltage $V_{xx}^{2\omega}$ at a fixed current density of  $j_{ac} = 1.5 \times 10^{10}$ A/m$^2$ with an applied field of $\mu_{0}$H = 400 mT. The curve is a fit to $V_{xx}^{2\omega}(0)\sin(\varphi)$. (c) Angular dependence of second harmonic transverse voltage $V_{xy}^{2\omega}$ at the same current density,  $j_{ac} = 1.5 \times 10^{10}$ A/m$^2$. The curve is a fit to $V_{xy}^{2\omega}(0)\cos(\varphi)$.}
\end{figure}

The samples we studied consist of a 20 nm thick epitaxial YIG film grown on a gadolinium gallium garnet ($\rm{Gd_3Ga_5O_{12}}$) substrate by RF sputtering~\cite{Mingzhong} and a 5 nm thick Pt film grown by DC sputtering in separate deposition systems. The YIG film is transferred in air and $Ar^{+}$ plasma cleaning is performed prior to the deposition of the Pt film. A Hall bar with a width of 4 $\mu$m and a length between the voltage contacts of 90 $\mu$m is fabricated using e-beam lithography and ion milling. The current flows in the x-direction and the voltage is measured both along the current direction (V$_{xx}$) and transverse to the current direction (V$_{xy}$) with separate electrical contacts (Fig. 1(a)). Lock-in amplifiers are used to measure the first harmonic and second harmonic voltages with phases $\mathrm{\phi _1 = 0^o}$ and $\mathrm{\phi _2 = -90^\circ}$ and a time constant of 300 ms. The AC current frequency is 953 Hz and its rms amplitude is indicated in the figures. All the angular dependent data are averaged 50 times to improve the signal-to-noise ratio. The measurements are conducted at room temperature.

Figure 1(b) and (c) show the second harmonic longitudinal $V_{xx}^{2\omega}$ and transverse $V_{xy}^{2\omega}$ voltage, respectively, as a function of the in-plane angle of a 400 mT magnetic field, a field sufficient to saturate the magnetization of the YIG layer. To confirm that the second harmonic signal is associated with the SSE, measurements were repeated as a function of the applied magnetic field magnitude~\cite{Vlietstra, Hayashi}. (See the Supplementary section~\cite{suppl}.) It is important to note that there are contributions to the second harmonic signal from the damping-like (DL) torque, field-like (FL) torque and Oersted (Oe) fields. By characterizing the field dependence of the second harmonic response these effects can be separated, particularly at small applied fields at which these torques and Oersted fields introduce additional structure in the angular dependence of the second harmonic signal. This is discussed in the supplementary section~\cite{suppl}, where the relative contributions of SSE, DL, FL and Oe field torques are determined~\cite{Vlietstra, Hayashi, avci}.  We find that for an applied field of 400 mT, the second harmonic signal is dominated by the SSE.

Figure 2(a) and (b) show the field dependence of second harmonic $V_{xy}^{2\omega}$ (Fig.~2(a)) and first harmonic $V_{xy}^{\omega}$ (Fig.~2(b)) response with the field applied along the current direction ($\varphi=0^\circ$) at a fixed AC current density of $1.5 \times 10^{10}$ A/m$^{2}$ as the DC component of the current density is varied, $j_{dc} = 0, \pm 2.5, \pm 5.0 \times 10^{10}$ A/m$^2$. The SSE response is expected to change sign when the magnetization direction reverses, which is evident in Fig.~2(a) in the step change in $V_{xy}^{2\omega}$ near zero field at the coercivity of the YIG ($\mu_0 H_c \simeq 10$ mT). The step in voltage in the second harmonic signal is nearly independent of the DC current. Interestingly, the first harmonic response depends systematically on the DC current. At zero DC current there is virtually no response, only small signal variations near zero field. However, when the DC current density is non-zero, a clear voltage step is evident near zero field, with a change in voltage that depends on the DC current. 

The magnitude of the first harmonic signal is about one order of magnitude larger than the second harmonic signal. In addition, the step in the first harmonic signal changes sign when the DC current is reversed. Figure~3(a) and (b) show how the steps in voltage depends on DC current. The step in the second harmonic signal $\Delta V_{xy}^{2\omega}$ (Fig.~3(a)) is slightly modified due to DC current, whereas there is a clear linear relation between the step in the first harmonic signal $\Delta V_{xy}^{\omega}$ (Fig.~3(b)) and the DC current.

\begin{figure}
\includegraphics[width=0.5\textwidth]{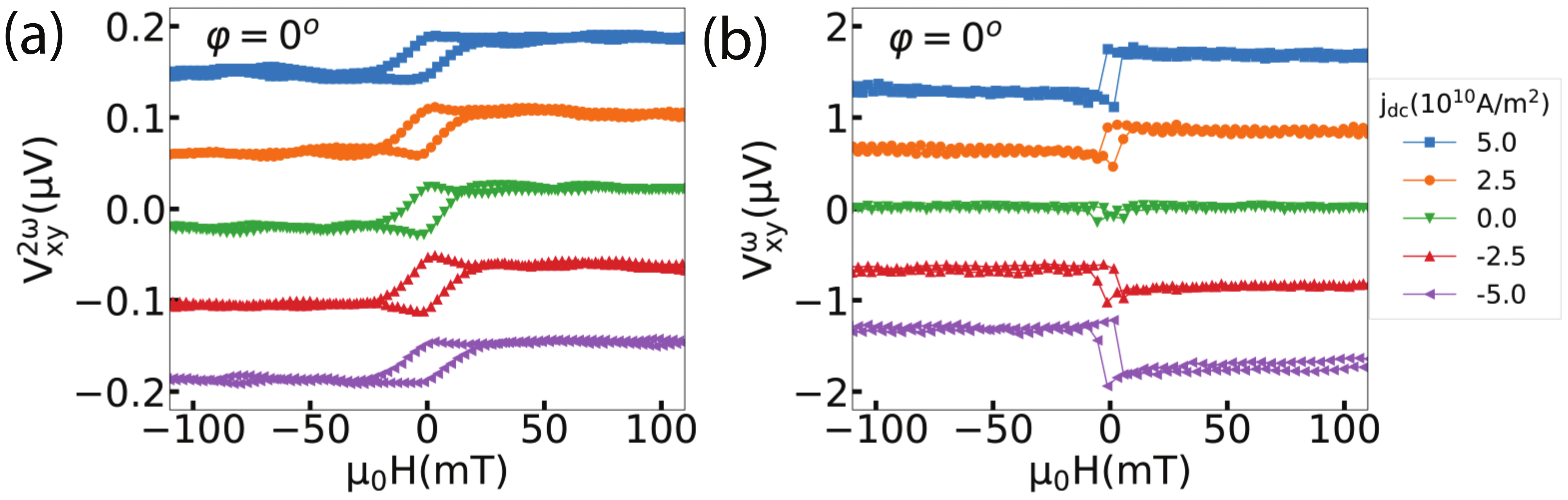}
\caption{Field dependent measurement of second and first harmonic transverse voltage with fixed AC current $j_{ac} = 1.5 \times 10^{10}$ A/m$^{2}$ and varying DC current. (a) Field dependence of $V_{xy}^{2\omega}$ at $\varphi = 0^\circ$ and $j_{dc} = 0, \pm 2.5, \pm 5.0 \times 10^{10}$ A/m$^{2}$. (b) Field dependence of $V_{xy}^{\omega}$ with $\varphi = 0^\circ$ and $j_{dc} = 0, \pm 2.5, \pm 5.0 \times 10^{10}$ A/m$^{2}$.}
\end{figure}

In order to understand this behavior one needs to consider Joule heating by the AC and DC current through the Pt. This leads to a power dissipation given by:
\begin{equation}
\begin{split}
&\quad P = [\sqrt 2 j_{ac}\cos({\omega}t) + j_{dc}]^2 R A^2 \\
&\quad \quad = [j_{ac}^2 cos(2\omega t) + 2\sqrt 2 j_{ac}j_{dc} cos(\omega t) + j_{ac}^2 + j_{dc}^2]RA^2,
\end{split}
\end{equation}
where $j_{ac}$ is the rms AC current density, $R$ is the resistance of the Pt and $A$ its cross sectional area, the film thickness times the width of the current line. The temperature gradient $\nabla T_z$ is proportional to the power dissipation. It follows that the SSE voltage generated has the following form:
\begin{equation}
V_{ISHE} \propto j_{ac}^2 cos(2\omega t) + 2\sqrt 2 j_{ac}j_{dc} cos(\omega t) + j_{ac}^2 + j_{dc}^2.
\label{Eq:VISHE}
\end{equation}
There is thus an SSE response at two times the oscillation frequency of the current, the second harmonic, $2\omega$, as expected, as well as a signal at frequency, $\omega$, the first harmonic. Thus the combination of AC  and DC currents provides a technique to measure the SSE voltage as a first harmonic response. The relative magnitude of the first and the second harmonic signals is given by $V_{xy}^{\omega}/V_{xy}^{2 \omega} = 2 \sqrt 2 j_{dc}/j_{ac}$. The first harmonic response is thus about 2.8 times larger than the second harmonic response when the AC and DC currents are the same. The linear relation between $\Delta V_{xy}^{\omega}$ and $j_{dc}$ in Fig.~3(b) confirms this model.

\begin{figure}[t]
\includegraphics[width=0.45\textwidth]{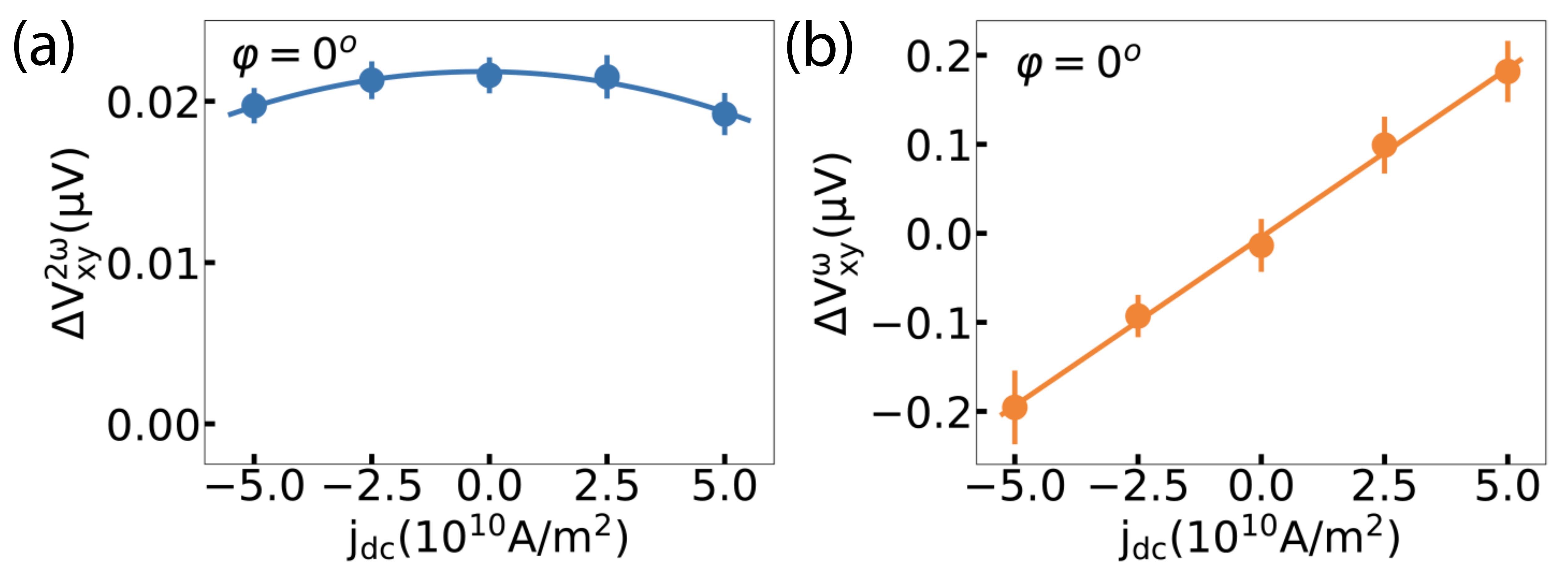}
\caption{Dependence of the second and first harmonic transverse voltage amplitudes on the DC current density with $j_{ac}$ fixed at $1.5 \times 10^{10}$ A/m$^2$. (a) Second harmonic transverse voltage versus DC current. (b) First harmonic transverse voltage versus DC current.}
\end{figure}

Further, we experimentally verify the symmetry and magnitude of the SSE first harmonic response in comparison to the conventional second harmonic signal. We have performed field dependent measurements of the first harmonic transverse voltage by sweeping the external magnetic field between -400 mT and +400 mT at different in-plane angles $\varphi$ from $0^\circ$ to $360^\circ$ at fixed  $j_{ac} = 1.5 \times 10^{10}$A/m$^2$ and $j_{dc} = \pm 5.0 \times 10^{10}$ A/m$^2$. Using these results, we have determined $\Delta V_{xy}^{\omega}$ using the procedure mentioned in the preceding section and plot its variation with $\varphi$ (Fig.~4). The SSE voltage is proportional to the projection of the magnetization on the axis perpendicular to the voltage probes. The temperature gradient is along the z-axis, whereas  the spin polarization is along the YIG magnetization direction. Therefore  the angular dependence of the first harmonic and second harmonic transverse response are $V_{xy}^{\omega} \propto 2\sqrt 2 j_{ac} j_{dc}\cos(\varphi), V_{xy}^{2\omega} \propto  j_{ac}^2\cos(\varphi)$, as seen experimentally. Measurements of $\Delta V_{xy}^{\omega}$ can be fitted well with $\Delta V_{xy}^{\omega}(0) \cos \varphi$, denoted by the solid line in Fig.~4. The second harmonic transverse voltage was measured with varying $\varphi$ at a fixed field of +400 mT and a fixed $j_{ac}$ $=$ $ 1.5 \times 10^{10}$ A/m$^2$(Fig.~1(c)). Equation~\ref{Eq:VISHE} predicts that for  $j_{ac} = 1.5 \times 10^{10}$A/m$^2$ and $j_{dc} = \pm 5.0 \times 10^{10}$ A/m$^2$, the relative magnitudes of the first and second harmonic signals should be $9.4$.
We have extracted the maximum $\Delta V_{xy}^{\omega}(0)=0.187 \pm 0.053 \; \mu$V and $\Delta V_{xy}^{2\omega}(0)=0.0227 \pm 0.0007  \;\mu$V by fitting the data in Fig.~4 and Fig.~1(d) respectively. The experimentally obtained ratio of the first and second harmonic signals is $ 8.2 \pm 2.7$. The experimentally obtained ratio is thus consistent with our simple AC and DC current heating model. The data in Fig.~4 clearly indicates that the first harmonic response has a much higher signal-to-noise ratio than that of the second harmonic voltage (Fig.~1(c)).

\begin{figure}[h]
\vspace{0.5 cm}
\includegraphics[width=0.33\textwidth]{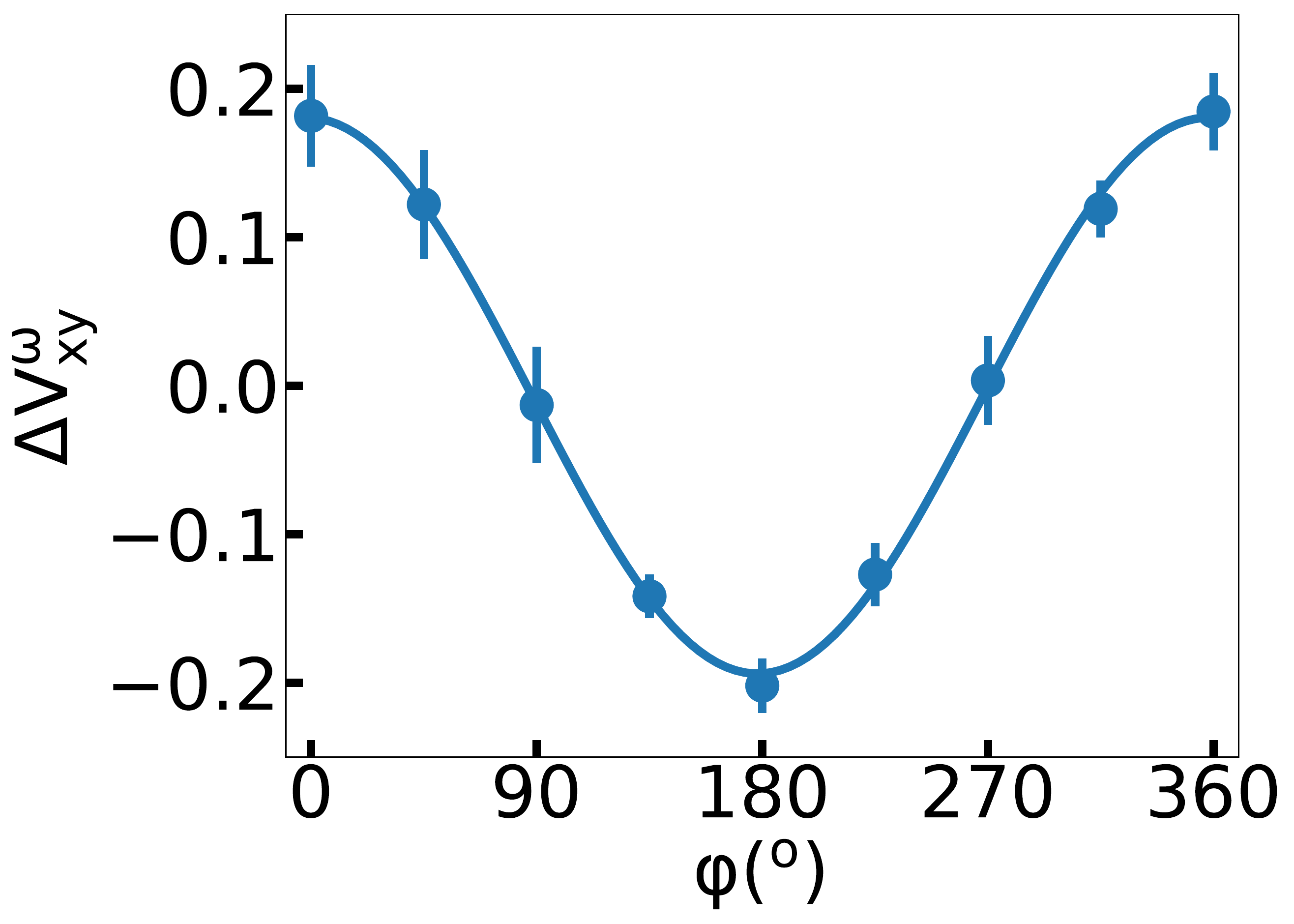}
\caption{Angular dependence of $\Delta V_{xy} ^{\omega}$ measured with an AC current of $1.5 \times 10^{10}$ A/m$^2$ and DC current $5.0 \times 10^{10}$ A/m$^2$. The curve is a fit to the data of the form $\Delta V_{xy} ^{\omega}(0)cos(\varphi)$.}
\end{figure}

In summary, we have determined the SSE-produced linear and nonlinear voltage responses in a YIG/Pt bilayer system. The second harmonic longitudinal voltage has a sine relation with respect to the in-plane field angle $\varphi$ when the YIG is saturated. Angular dependence measurement of the longitudinal and transverse voltages as a function of the applied field magnitude enabled estimation of the contributions from SSE, DL, FL and Oe field torques. It was found that the SSE dominates over the other contributions when the applied field is sufficient to saturate the YIG layer. In addition, by applying an AC current with DC bias, we determined that SSE can be measured by a first harmonic lock-in technique, and can be more sensitive and have higher signal to noise than the conventional second harmonic metthod. This technique can be used to characterize the SSE in ferromagnetic (or ferrimagnetic) and non-magnetic bilayer systems as well as to study nonlinear thermoelectric effects and spin dynamics induced by temperature gradients.

\section*{Acknowledgements}
The instrumentation used in this research was support in part by the Gordon and Betty Moore FoundationÕs EPiQS Initiative through Grant GBMF4838 and in part by the  National Science Foundation under award NSF-DMR-1531664. This work was supported partially by the MRSEC Program of the National Science Foundation under Award Number DMR-1420073. ADK received support from the National Science Foundation under Grant No. DMR-1610416.
At CSU, film growth was supported by the U.S. National Science Foundation (EFMA-1641989), and film characterization was supported by the U.S. Department of Energy, Office of Science, Basic Energy Sciences (DE-SC-0018994).

\section{Supplemental materials: First harmonic measurements of the spin Seebeck effect}
\subsection*{Separation of the Spin Seebeck Effect, anti-damping spin-orbit torque, field-like torque and Oersted field contributions}
In the main text we state that the origin for both the longitudinal and transverse second-harmonic signals for a 400 mT applied field are due to the spin Seebeck effect arising from current-induced Joule heating in the Pt strip. Here, we estimate the relative contributions to $V_{xy}^{2\omega}$ and $V_{xx}^{2\omega}$ from SSE, anti-damping spin-orbit torque (AD), field-like torque (FL) and Oersted field contributions (Oe). 

Angular dependence measurements with applied magnetic field ranging from 2 mT to 400 mT are used to separate there contributions using the following relations~[S1,S2]:
\begin{equation} \label{eq:s1}\tag{S1}
\begin{split}
\quad V_{xx}^{2\omega}(\varphi)\ &= \Delta V_{xx}^{2\omega, \varphi} \sin(\varphi + \Delta \varphi)  \\
&+  \Delta V_{xx}^{2\omega, 3\varphi} \sin(\varphi + \Delta \varphi) \cos^2(\varphi + \Delta \varphi) \\ &+ A_0 
\end{split}
\end{equation}
$\Delta V_{xx}^{2\omega, \varphi}$  is the SSE and AD contributions and $\Delta V_{xx}^{2\omega, 3\varphi}$ is the FL and Oe contributions. $A_0$ is the offset of the signal.

\begin{equation} \label{eq:s2}\tag{S2}
\begin{split}
\quad  V_{xy}^{2\omega}(\varphi) &=  \Delta V_{xy}^{2\omega, \varphi} \cos(\varphi + \Delta \varphi) \\
&+  \Delta V_{xy}^{2\omega, 3\varphi} \cos(\varphi + \Delta \varphi)\cos[2(\varphi + \Delta \varphi) ]\\
& + B_0 
\end{split}
\end{equation}
$\Delta V_{xy}^{2\omega, \varphi}$  is the SSE and AD contributions and $\Delta V_{xy}^{2\omega, 3\varphi}$ is the FL and Oe contributions. $B_0$ is the offset of the signal.

We plot $\Delta V_{xx/xy}^{2\omega, \varphi}$ versus the inverse applied field, $\mu _0 H$:
\begin{equation} \label{eq:s3}\tag{S3}
\quad  \Delta V_{xx/xy}^{2\omega, \varphi} = \frac{k_{xx, \varphi} }{ \mu _0 H}+ C_{xx/xy}^{\varphi} \\
\end{equation}
to determine the slope $k_{xx, \varphi} $ and intercept $C_{xx/xy}$.

We then plot $\Delta V_{xx/xy}^{2\omega, 3\varphi}$ versus the inverse applied field, $\mu _0 H$:
\begin{equation} \label{eq:s4}\tag{S4}
\begin{split}
& \quad  \Delta V_{xx/xy}^{2\omega,3\varphi} = \frac{k_{xx, 3\varphi} }{ \mu _0 H} + D_{xx/xy} ^{3\varphi}\\
\end{split}
\end{equation}
to determine the slope $k_{xx, 3\varphi}$and intercept $D_{xx/xy}$.


\begin{figure}[b]
\includegraphics[width=0.45\textwidth]{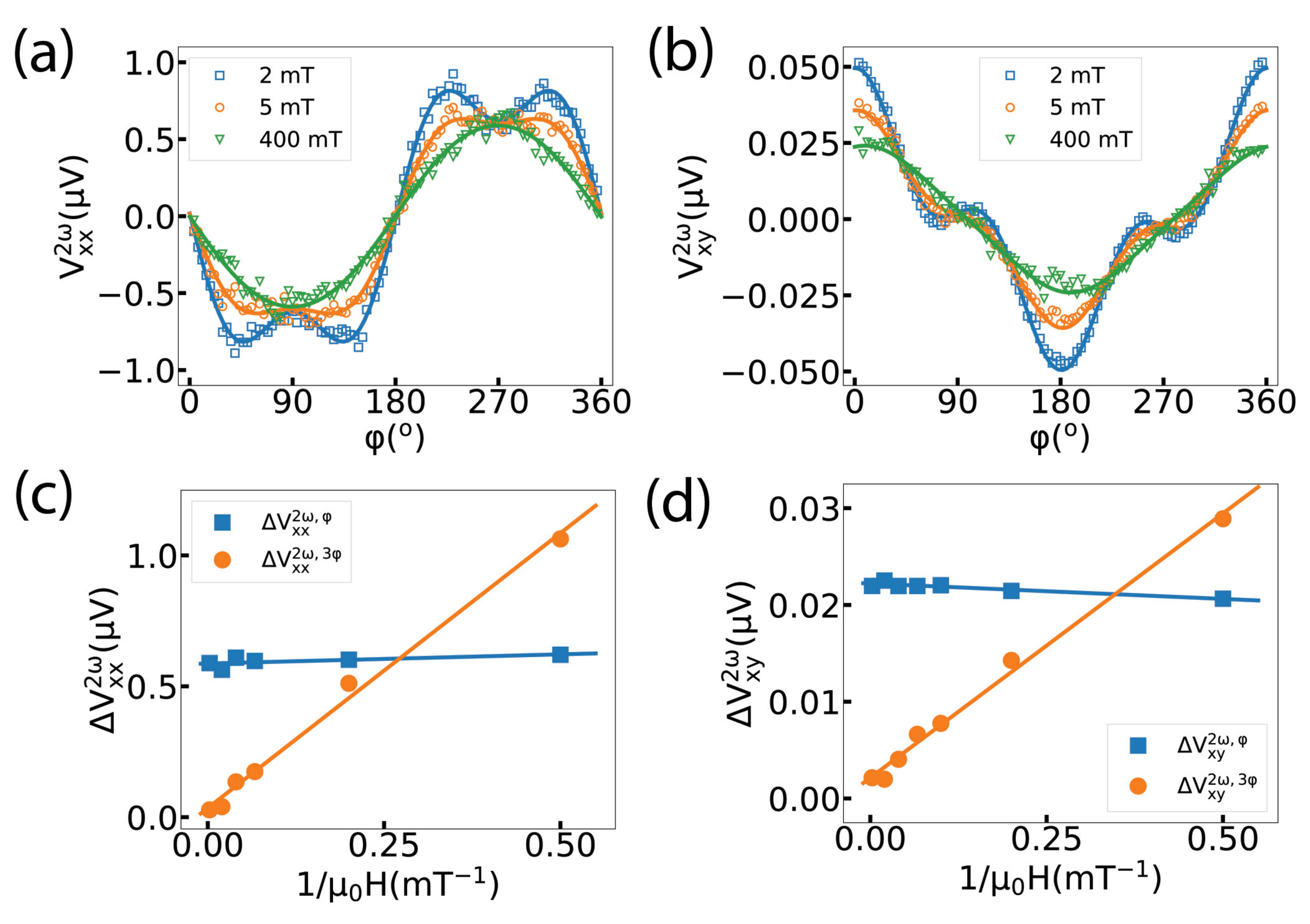}
\setcounter{figure}{0}
\renewcommand{\figurename}{Fig. S}
\caption{Second harmonic measurements of $V_{xx}^{2\omega}$ and $V_{xy}^{2\omega}$ with different applied magnetic fields. The AC current density $j_{ac} = 1.5 \times 10^{10} A /m^{2} $ and applied magnetic field $\mu _0 H$ ranges from 2 to 400 mT. Solid lines denotes the fits.
(a) Angular dependence of $V_{xx}^{2\omega}$, data fits to equation (\ref{eq:s1});
(b) Angular dependence of $V_{xy}^{2\omega}$, data fits to equation (\ref{eq:s2});
(c) Field dependence of $\Delta V_{xx}^{2\omega, \varphi}$ and $\Delta V_{xx}^{2\omega, 3\varphi}$ with the intercepts $C_{xx}^{\varphi}$ = $0.59\ \mu V$ and $ D_{xx}^{\varphi}= 3.8 \times 10^{-3}\ \mu V$;
(d) Field dependence of $\Delta V_{xy}^{2\omega, \varphi}$ and $\Delta V_{xy}^{2\omega, 3\varphi}$ with the intercepts $D_{xy}^{\varphi}$ = $22.2 \times 10^{-3}\ \mu V$ and $ D_{xx}^{\varphi}= 2.1 \times 10^{-3}\ \mu V$.
}
 \end{figure}

Fig.~S1 (a) and Fig.~S1 (b) show the angular dependence of $V_{xx}^{2\omega}$ and $V_{xy}^{2\omega}$ versus applied magnetic field ranging from 2 to 400 mT. When $\mu _0 H = 400$ mT, $V_{xx}^{2\omega}$ and $V_{xy}^{2\omega}$ fit well to  $\sin(\varphi)$ and $\cos(\varphi)$, with negligible $3\varphi$-contributions. For $\mu _0 H$ smaller than 25 mT, clear $3\varphi$-symmetry can be observed due to the non-negligible FL + Oe effects, indicating that the applied field torque is comparable to the AD, FL and Oe torques. By fitting $V_{xx}^{2\omega}$ with Eqn.~\ref{eq:s1} and $V_{xy}^{2\omega}$ with Eqn.~\ref{eq:s2}, we can extracted the relative SSE, AD, FL and Oe contributions. $\Delta V_{xx}^{2\omega, \varphi}$ and $\Delta V_{xy}^{2\omega, \varphi}$ are denoted as the $\varphi$-contributions while $\Delta V_{xx}^{2\omega, 3\varphi}$ and $\Delta V_{xy}^{2\omega, 3\varphi}$ indicate the $3\varphi$-contributions. Fig.~S1(c) and Fig.~S1(d) show the field dependence of $\varphi$ and $3\varphi$-contributions. The $\varphi$ contributions are almost independent of the applied magnetic fields, showing negligible AD. However, the $3\varphi$ contributions are inversely proportional to the applied magnetic fields with negligible intercepts. When the applied field is in the range of $\sim$3 mT, the $3\varphi$ contributions surpass the $\varphi$ contributions, showing increasing FL and Oe effects at low fields.

\end{document}